\documentstyle[prd,preprint,aps]{revtex}
\tightenlines
\newcommand{\be}{\begin{equation}}
\newcommand{\ee}{\end{equation}}
\newcommand{\ba}{\begin{eqnarray}}
\newcommand{\ea}{\end{eqnarray}}
\newcommand{\oh}{\displaystyle{\frac{1}{2}}}
\begin{document}
\draft
\title{Induced Parity Breaking Term in Arbitrary Odd Dimensions
at Finite
Temperature}
\author{C.D.~Fosco$^a$\thanks{CONICET, Argentina}\,,
G.L.~Rossini$^{b\, *}$\, and\,
F.A.~Schaposnik$^b$\thanks{Investigador CICBA, Argentina}
\\
{\normalsize\it $^a$Centro At\'omico Bariloche, 8400 Bariloche,
Argentina}\\
{\normalsize\it $^b$Departamento de F\'\i sica, Universidad
Nacional de La
Plata}\\ {\normalsize\it C.C. 67, 1900 La Plata, Argentina}}

\maketitle
\begin{abstract}

We calculate the exact parity odd part of the effective action
($\Gamma_{odd}^{2d+1}$) for massive Dirac fermions in  $2d+1$
dimensions at
finite temperature, for a certain class of gauge field
configurations. We
consider first Abelian external gauge fields, and then we deal with
the case
of a  non-Abelian gauge group containing an Abelian $U(1)$
subgroup. For both
cases, it is possible to show that the result depends on
topological
invariants of the gauge field configurations, and that the gauge
transformation properties of $\Gamma_{odd}^{2d+1}$ depend only on those
invariants and on the winding number of the gauge transformation.

\end{abstract}

~


\bigskip

\newpage

\section{INTRODUCTION}

The issue of parity breaking at finite temperature in $3$ dimensional
gauge theories with massive fermions posed a puzzle concerning the induced
effective action: perturbative calculations indicated that it was simply a
Chern-Simons (C-S) term times a coefficient that was a smoothly varying
function of the temperature but this was in contradiction with gauge
invariance \cite{pis}-\cite{cfrs}.

A crucial advance was made in \cite{dll} by studying a $D=1$ solvable model
for which the {\it exact} effective action was gauge invariant although
perturbative expansions produced gauge-noninvariant results. Subsequently, 
it was shown that the same phenomenon also takes place in $2+1$ dimensions. 
This was proven through non-perturbative calculations of the effective 
action in the Abelian case \cite{dgs} and of its explicit, exact, 
temperature dependent parity breaking part both in the Abelian and 
non-Abelian cases \cite{frs},  for particular gauge backgrounds. 

These results were discussed in connection with reduction of
C-S terms by a symmetry \cite{JP1}-\cite{JP2} and also
confirmed by several alternative calculations
\cite{AF}-\cite{P}.

It is the purpose of the present work to extend the results in \cite{frs} to
the case of {\it arbitrary} odd dimensions, $D = 2d +1$. Indeed, of the many
interesting properties enjoyed by odd dimensional  quantum field theories,
not
the less important is the possibility of  equipping a gauge field with a
Chern-Simons action. This parity breaking object is invariant under gauge
transformations connected to the identity, but not necessarily so for "large"
ones. Demanding invariance of the partition function under large gauge
transformations has important consequences, particularly for the cases of
non-trivial spacetimes (even in the  Abelian case), or when the gauge group
is
non-Abelian.

As in the $D=3$ case, for arbitrary $D = 2d + 1$ dimensions the
C-S action arises as the result of integrating out fermionic degrees of
freedom at zero temperature. At finite $T$, temperature dependent parity
breaking terms are also induced by integrating fermionic degrees of freedom, in
such a form that their zero $T$ limit coincides, as we shall see,
with the C-S action \cite{J}-\cite{ADM}.

The clue in the approach of \cite{frs} to the $3$ dimensional case was to
choose a particular gauge background in which the temperature dependence in 
the parity breaking part of the effective action can be factored out, 
leaving all the spatial information encoded in the form of the  $2$ 
dimensional chiral anomaly. The main point in the present paper is to show  
that the same holds in $D = 2d + 1$ dimensions. Namely,  
for particular gauge 
field backgrounds, the temperature dependence is isolated in a factor that 
can be related to the Polyakov loop and the spatial components of the gauge 
configuration give rise to a factor which is nothing but the chiral 
anomaly, now in $2d$ dimensions. This is done both in the Abelian case
(Section II)  and in the
non-Abelian one (Section III) for a particular choice of the 
gauge background which is, 
however, sufficiently general as to allow to infer qualitative properties in 
the general case.  We give a summary and discussion of our results
in section IV.

\section{THE ABELIAN CASE}

Let us start by stressing that in the imaginary time formalism of finite
temperature Quantum Field Theory, the effective action for $D=2d+1$
dimensional Dirac fermions with mass $M$ can be a non-extensive quantity
whose
temperature dependent, parity-odd part,
 will be called here $\Gamma_{odd}^{2d+1} (A,M)$.  It is
defined as
\be
\Gamma_{odd}^{2d+1} (A,M)  \;=\;\oh (\Gamma ^{2d+1} (A,M)-\Gamma
^{2d+1} (A,-M))
\label{gammaodd}
\ee
where
\be
\exp\left( {- \Gamma^{2d+1}  (A,M)} \right)\;=\;  \int {\cal D} \psi \,
{\cal D} {\bar \psi} \; \exp \left[ - S_F (A, M) \right] ,
\label{ealagamma}
\ee
and the  Euclidean action $S_F (A,M)$ is given by
\be
S_F (A,M) \;=\; \int_0^\beta d \tau \int d^{2d} x \; {\bar \psi} (
\not \!
\partial + i e \not \!\! A + M ) \psi \; .
\ee
The Euclidean $\gamma$ matrices in $2d+1$ dimensions are denoted as
$\gamma_o, \gamma_1 , \cdots, \gamma_{2d}$. From the point of view of the 
$2d$ dimensional theories (with coordinates ${x}$) that will arise below, 
$\gamma_0$ will act as a $\gamma_5$ chirality matrix. 

The fermionic fields in (\ref{ealagamma}) obey  antiperiodic boundary
conditions in the timelike direction
\be
\psi (\beta , {x}) \;=\; - \, \psi (0 , {x}) \;\;\;\;,\;\;\;\;
{\bar \psi} (\beta,{x}) \;=\; - {\bar \psi} (0, {x}) \;\;, \forall {x}
\; .
\label{fbc}
\ee
The gauge field, instead, is periodic in the same direction
\be
A_\mu (\beta,{x}) \;=\; A_\mu (0,{x}) \;\;,\;\; \forall {x} \; .
\label{gbc}
\ee

The effective action $\Gamma^{2d+1}  (A,M)$ is, as usual, written in terms 
of a fermionic determinant 
\be
\Gamma^{2d+1}  (A,M)= -\log \det ( \not \! \partial + i e \not \!\! A \,+\, 
M ) \;.\label{fdet1} 
\ee

In order to get an exact result we choose a particular gauge field
background
which corresponds to a vanishing electric field and a time-independent
magnetic field,
\be
A_0 = A_0(\tau),
\label{0}
\ee
\be
A_j = A_j({x})\,\,\, (j=1,2,\cdots, 2d)
\label{1}
\ee
or any equivalent configuration obtained from this by a gauge
transformation.

Using the same arguments as in ref \cite{frs} we can always perform a (non
anomalous) gauge transformation of the fermionic fields in the functional
integral defining the fermionic determinant in eq.(\ref{ealagamma}), so that
as
the zero component of the gauge field becomes a constant that will be called
$\tilde A_0$,
\be
{\tilde A}_0 \;=\; \frac{1}{\beta} \, \int_0^\beta \, d \tau \, A_0 (\tau)
\; .
\ee

After redefining the fermionic fields according to this prescription, we
get
\be
S_F ( A_j , {\tilde A}_0 , M) \;=\; \int_0^\beta d \tau \int d^{2d} x \; {\bar
\psi}
( \not \! \partial + i e ( \gamma_j A_j + \gamma_0 {\tilde A}_0 ) + M )
\psi \; ,
\ee
where there is now no explicit $\tau$ dependence in
the background. Then, if one  performs a Fourier
transformation on the time variable for $\psi$  and ${\bar
\psi}$
\ba
\psi (\tau, {x}) &=& \frac{1}{\sqrt \beta} \,
\sum_{n=-\infty}^{+\infty} \,
e^{i \omega_n \tau} \psi_n ({x}) \nonumber\\ {\bar \psi} (\tau, {x}) &=&
\frac{1}{\sqrt \beta} \, \sum_{n=-\infty}^{+\infty} \, e^{-i
\omega_n \tau} {\bar
\psi}_n ({x}) \;,
\label{Four}
\ea
where $\omega_n = (2 n +1) \frac{\pi}{\beta}$ is the usual
Matsubara frequency
for fermions, the Euclidean action becomes an infinite series of
decoupled
$2d$ dimensional actions, one for each Matsubara mode:
\be
S_F ( A_j , {\tilde A}_0 , M) \;=\;
\sum_{n=-\infty}^{+\infty}
\int d^{2d} {x} {\bar \psi}_n ({x}) \left[ \not \! d \,+\, M \,+\, i
\gamma_0
(\omega_n + e {\tilde A}_0) \right] \psi_n ({x}).
\ee
Here, $\not \! d$ is the $2d$ Euclidean Dirac operator
corresponding to the
spatial coordinates and the spatial components of the gauge field
\be
\not \! d \;=\;\gamma_j (\partial_j + i e A_j).
\label{sup}
\ee

As the Matsubara modes introduced in eq.(\ref{Four}) are decoupled,  the
$2d+1$ determinant arising from
fermion integration becomes an infinite product of the determinants
of $2d$
Euclidean Dirac operators
\be
\det ( \not \! \partial + i e \not \! A \,+\, M )_{2d +1} \; = \;
\prod_{n=-\infty}^{n=+\infty} \det [\not \! d + M + i \gamma_0 (\omega_n
+ e {\tilde A}_0) ]_{2d} \;.
\ee

Remarkably, the parity odd piece of $\Gamma$ 
defined by eqs.(\ref{gammaodd}),(\ref{fdet1}) can then  be factorized, 
for arbitrary  odd space-time dimension
$D$, following  the
procedure discussed in ref.\ \cite{frs} for $D=3$. Indeed, for
any given mode, one can factorize the $2d$ determinant into mass even and
mass
odd pieces through a chiral transformation as
\be
\det [\not \! d + M + i \gamma_0 (\omega_n
+ e {\tilde A}_0) ]_{2d} \;=\; J_n[A,M] \; \det [\not \! d +  \rho_n ]_{2d}
\label{fuji1}
\ee
where
\be
\rho_n \;=\; \sqrt{ M^2 + ( \omega_n + e {\tilde A}_0 )^2 }\;
\ee
and $J_n[A,M]$ is the anomalous Jacobian of a chiral transformation  in $2d$
dimensions,
\be
\psi_n (x) \;=\; \exp\left(
{- i \frac{\phi_n}{2} \gamma_0}\right) {\psi'}_n (x) \;\;,\;\;
{\bar \psi}_n (x) \;=\; {{\bar \psi}'}_n (x) \exp\left(
{- i \frac{\phi_n}{2}
\gamma_0} \right) ,
\label{quiral}
\ee
with phase
\be
\phi_n \;=\; {\rm arctan} ( \frac{\omega_n + e {\tilde A}_0}{M} ) \; .
\ee

It is important to stress at this point
the reason why this procedure can be pursued in more
than $3$ dimensions. The Jacobian for a constant chiral transformation is
exactly known for {\em any} even dimension, and this is all we need to
evaluate the parity-odd part of the effective
action. In fact, from definition
(\ref{gammaodd}),(\ref{fdet1}) we see that $\Gamma_{odd}$ depends solely on the
Jacobians $J_n$,
\be
\Gamma_{odd} \;=\; - \sum_{n=-\infty}^{n=+\infty} \, \log J_n[A,M].
\ee
Now, each Fujikawa Jacobian can be seen to give
\be
J_n[A,M] \;=\; \exp ( -i{\phi_n}  \int d^{2d}x {\cal A}_{2d}[A] )\;,
\label{J}
\ee
with ${\cal A}_{2d}[A]$ denoting the $2d$ dimensional chiral
anomaly. Then
\be
\Gamma_{odd}^{2d+1} \;=\; i
  \, \Phi  \; \int d^2 x {\cal A}_{2d}[A]\; ,
\ee
where
\be
\Phi= \sum_{n=-\infty}^{n=+\infty} \phi_n.
\label{sumaw}
\ee
Note that the phases $\phi_n$  contain at this stage {\it all} the 
dependence on the Matsubara frecuencies $\omega_n$. Moreover, they are {\em 
independent} of the number of spacetime dimensions, and
hence the sum over  
$\phi_n$ is the same as the one already calculated   in ref.\ \cite{frs}
for the $D=3$ case, 
\be
\Phi \;=\; {\rm arctan} \left[
\tanh(\frac{\beta M}{2}) \tan ( \oh e \beta {\tilde A}_0 ) \right] .
\label{suma1}
\ee
Thus the parity-odd part of $\Gamma$
finally reads
\be
\Gamma_{odd}^{2d+1} \;=\; i \;
{\rm arctan} \left[ \tanh(\frac{\beta M}{2})
\tan ( \frac{e}{2} \int_0^\beta d \tau
A_0(\tau) ) \right]\, \int d^{2d} x {\cal A}_{2d}[A] \; .
\label{espl'}
\ee

This is one of the main results in our paper. We have been able to
compute the {\it exact} temperature dependent parity-odd piece of the effective
action for massive fermions in a gauge-field background {\it in arbitrary
dimensions}. Remarkably, the temperature dependent factor is universal
in the sense it does not depend on the number $D=2d+1$ of
the space-time dimensions. The particular background we have chosen
makes evident the role of Polyakov loop in the temperature dependent
factor of the effective action as will be discussed below.
Concerning   the
dependence on the spatial components of the gauge field, it is
just given by the $2d$ chiral anomaly. As we shall see
below, all these
feautures are also valid in the non-Abelian case.

The $D = 2d+1 = 3$ case was discussed in detail in ref.\ \cite{frs}.
Let us then, 
as another example,  write down here the explicit expressions for
$D=2d+1=5$. In
this case the well known $4$ dimensional chiral anomaly is given by
\be
{\cal A}_4 = - \frac{e^2}{16\pi^2} {F_{ij}} ^*\!F_{ij},
\ee
so that
\be
\Gamma_{odd}^5 \;=\; -i\, {\rm arctan} \left[ \tanh(\frac{\beta M}{2}) \tan 
( \frac{e}{2} \int_0^\beta d \tau A_0(\tau) ) \right]\,\frac{e^2}{16\pi^2} 
\int d^{4} x {F_{ij}} ^*\!F_{ij} \; . \label{espl''} 
\ee
Let us note that
the anomaly factor in (\ref{espl''}) can be non trivial
in the Abelian case according
to the properties of  the $2d$ manifold ${\cal M}$ on which
${\cal A}_4$ is integrated. For example if
${\cal M} = S^2 \times S^2$, $\int {\cal A}_4 d^4x = 2$, the
smallest value it can take for spin manifolds (for non-spin
manifolds it can take also odd values \cite{wit}).

In arbitrary dimension $D=2d$, we quote the form of the Abelian chiral
anomaly, which is simply given by
\be
{\cal A}_{2d}=-\frac{(-e)^d}{(4\pi)^d} \frac{1}{d!}\epsilon_{\mu_1 \mu_2
\cdots
\mu_{2d}} F_{\mu_1 \mu_2} \cdots F_{\mu_{2d-1}\mu_{2d}}
\ee
(see for instance \cite{Ball}).

Let us check that the $D=5$ expression in eq.(\ref{espl''}) has the proper 
$T=0$ limit, i.e. it reduces to the usual Chern-Simons term. This limit 
reads 
\be
\Gamma_{odd}^5 \to -i \frac{M}{|M|}\frac{e^3}{32\pi^2}
 \int_0^\beta d \tau A_0(\tau) \, \int d^{4} x F_{ij}
{^*\!F_{ij}}.
\label{Gamma5}
\ee
This is exactly  the form taken by the Chern-Simons term
(with a coefficient that is half the value necessary
for making $\exp( S_{CS})$ gauge invariant even under
large gauge transformations),
\be
S_{CS}^5= \frac{ie^3}{48\pi^2}\int d\tau \,d^4x
\epsilon_{\mu\nu\rho\sigma\lambda}A_{\mu}\partial_{\nu}
A_{\rho}\partial_{\sigma}A_{\lambda} \label{CS5} 
\ee
when evaluated on the configurations restricted by eqs. (\ref{0}) and
(\ref{1}). It is interesting to note that in the $T \to 0$ limit
$\exp( -\Gamma^5_{odd})$ is nothing but the Polyakov loop with a
coefficient that corresponds to a topological
invariant for the $2d$-dimensional gauge theory.

Notice that, in order to take into account all contributions to the  above
mentioned configurations, one must approach them from a general  one. Some
terms that naively vanish for these configurations are actually  finite
since
the limit $A_0({x})\to constant$ is undetermined. The same kind of
undetermination is found in the $D=3$ case; we explain here the correct
procedure for that simplest example (see ref.\ \cite{frs}), since
complications arising in higher dimensions are unessential. We write
$S_{CS}^3$ in momentum space,
\be
S_{CS}^3=-\frac{e^2}{4\pi} \int
\frac{d^3p}{(2\pi)^3}\epsilon_{\mu\nu\lambda}A_{\mu}(-p)
p_{\nu}A_{\lambda}(p),
\ee
and explicitly separate the $\mu=0$ index,
\begin{eqnarray}
S_{CS}^3 & = & -\frac{e^2}{4\pi} \int
\frac{d^3p}{(2\pi)^3}\left[\epsilon_{jk}A_{0}(-p)
p_{j}A_{k}(p) + \epsilon_{jk}A_{j}(-p) p_{k}A_{0}(p) -\right. \nonumber\\
& & \left.\epsilon_{jk}A_{j}(-p)
p_{0}A_{k}(p)\right].
\end{eqnarray}
It is immediately seen that the first two terms contribute by the
same amount,
and that this amount is finite because $A_{k}(p)$ has a pole in
$p_j$; the
last term vanishes because $A_{k}(p)$ is also proportional to
$\delta(p_0)$
(see \cite{AF} for details).

In other words, a safe procedure in coordinate space is the
following: we
first write the integrand without spatial derivatives acting on
$A_0$, by
means of integrations by parts, and only then use the fact that
$\partial_0
A_j=0$. This gives twice the result of the naive restriction given
by using
$\partial_0 A_j=\partial_i A_0=0$.

The same check can be done in the general $2d+1$ dimensional case.
The correct
evaluation of the Chern-Simons term gives then $d+1$ times the
naive result.
In particular, one gets full agreement between eqs.\ (\ref{Gamma5}) and
(\ref{CS5}).

Let us end this section by discussing the issue of gauge invariance
under {\it large} gauge-transformations, a question which, as explained in
the introduction, was put in doubt by perturbative calculations
for $D=3$. For the particular abelian background we are considering,
such transformations $\Omega$ wind around the cyclic time direction,
$\Omega(\beta,x) = \Omega(0,x) + (2\pi/e)k$, with $k \in Z$.
The exact
result we have obtained for the temperature-dependent, 
parity-odd effective action, eq.(\ref{espl'}), shows that  large
gauge transformations may change the temperature dependent factor
if its winding number is odd. Indeed,  such a transformation,
say with a winding number $k = 2p+1$, shifts the argument of the
tangent in $(2p +1)\pi$. One has to keep track of this shift 
by shifting the branch used for the arctan definition. Now, 
if the integral of the anomaly is an even integer $n=2m$ the
total change of the effective action is $2m(2p+1)\pi i$ and
hence $\exp(-\Gamma_{odd})$ remains unchanged. In contrast,
if $n = 2m+1$, it changes its sign. However,
 as it is well known, there is a 
mass and temperature independent parity anomaly
contribution  which we have
not included in (\ref{espl'}) \cite{DJT}-\cite{GRS}
 which precisely changes its sign 
so that the exponential
of the complete effective action is indeed gauge-invariant.

\section{THE NON-ABELIAN CASE}

We extend here the analysis 
to the  a non-Abelian case  which,
for a
special class of gauge field configurations, generates a
$\Gamma_{odd}^{2d+1}$
with nice topological properties. The model is defined by its
Euclidean action
\be
S_F  \;=\; \int_0^\beta d \tau \int d^{2d} x \; {\bar \psi} ( \not
\! \!D + M
)
\psi \;
\ee
where now
\be
\not \!\! D_{\mu} = \partial_{\mu} + g A_{\mu}
\ee
and the antihermitean gauge connection $A_{\mu}$ corresponds to the Lie
algebra of some group $G$. For  computation simplicity
we shall consider that $G$
has   an Abelian $U(1)$
factor so that we can
decompose  $A_{\mu}$ as
\be
A_{\mu} = i A_{\mu}^0 + A_{\mu}^a \tau_a,
\ee
where $A_{\mu}^0$ is the component corresponding to  the Abelian factor
$U(1)$, while $A_{\mu}^a $ denotes the ones for the non-Abelian subgroup
that for definiteness we shall take to be $SU(N)$. The matrices $\tau_a$ 
are the generators for $SU(N)$, satisfying the relations 
\be
[\tau_a , \tau_b] \,=\,f_{abc} \,\tau_c \;\;\;
\tau_a^\dagger = - \tau_a, \;\;\;
tr(\tau_a \tau_b)=-\frac{\delta_{ab}}{2}.
\ee
We now fix the class of gauge configurations we consider to those
verifying
the conditions
\be
\begin{array}{ll}
A_0^0=A_0^0(\tau) ,& A_j^0=0,
\\ A_0^a=0, & A_j^a=A_j^a(x).
\end{array}
\ee
The $\tau$ dependence of $A_{\mu}$, present only through $A_0^0$, may be
eliminated by an Abelian gauge transformation just as in the Abelian case.
Then the fermionic action becomes
\be
S_F ( A_j^a , {\tilde A}_0^0 , M) \;=\; \int_0^\beta d \tau \int d^{2d} x 
\; {\bar \psi} ( \not \! \partial + g( \gamma_j A_j \tau_a + i\gamma_0 
{\tilde A}_0^0 ) + M ) \psi \; . 
\ee
Now, due to the commutativity of ${\tilde A}_0^0$ with $A_j^a$, the  same
steps leading to the calculation of $\Gamma_{odd}^{2d+1}$ may be  performed
here, with trivial modifications, except for the fact that the anomaly
${\cal A}$ will be the one corresponding to a $2d$ dimensional Abelian 
chiral rotation for a Dirac fermion in presence of a non-Abelian connection 
$A_j^a(x)$.  This gauge field is to be regarded as an arbitrary $SU(N)$ 
gauge field for the $2d$ dimensional sector of the theory. The anomaly is 
then  of course  the well known ``singlet'' anomaly \cite{Ball} 
\be
{\cal A}_{2d}= -
\frac{(ig)^d}{(4\pi)^d} \frac{1}{d!}\epsilon_{j_1 j_2
\cdots
j_{2d}} tr[F_{j_1 j_2} \cdots F_{j_{2d-1}j_{2d}}].
\ee
Now, as the integral of ${\cal A}_{2d}$ is proportional to the
Pontryagin index
 of the configuration
\be
\int  d^{2d}x  {\cal A}_{2d}(x) = n  
\ee
we may write $\Gamma_{odd}^{2d+1}$ as
\be
\Gamma_{odd}^{2d+1}= i \;
{\rm arctan} \left[ \tanh(\frac{\beta M}{2})
\tan ( \frac{g}{2} \int_0^\beta d \tau
A_0(\tau) ) \right] n.
\ee

Some remarks about this expression are in order. First, note that  it
is non-trivial
only
for $D=2d+1 >3$ dimensions since for
$2d = 2$   the
singlet anomaly vanishes.  Depending on the gauge group and the $2d$
manifold over which the anomaly is integrated  
the Pontryagin index $n$ can be a non trivial integer. 
Second, it is an object which is sensitive to  large gauge transformations 
in $2d+1$ spacetime, putting together the winding 
associated with the 
timelike direction $\tau$ (reflected in $A_0^0$),  with the usual winding 
transformations in $2d$. However, the restrictions on the background gauge 
fields that we have imposed do not allow us to analyse general large 
gauge transformations although we expect that, as in the Abelian
case, gauge-invariance is respected. 

\section{SUMMARY AND DISCUSSION}

As a summary of  our results we should like
to stress the following points

\noindent (i) The exact finite temperature
effective action induced by massive fermions in arbitrary odd dimensions
has the proper behavior under gauge transformations. Although
finite temperature calculations to fixed perturbative order 
necessarily violate gauge invariance, when all orders are taken into
account the invariance is restored.

\noindent (ii) Using a certain class of gauge field configurations, the
temperature dependent parity-odd part of the effective action
(i.e., the relevant part to investigate possible gauge invariant
violations at finite temperature) can be calculated {\it exactly
in arbitrary odd dimensions}.
The result is a gauge invariant action which is not just a 
Chern-Simons term with a temperature-dependent coefficient but
which reduces, in the low temperature regime, to this product and
confirms, at $T=0$ that massive fermions induce a C-S action.

\noindent (iii) An exact calculation was possible
because $\gamma_0$ in $2d +1 $ dimensions can be always taken as
the chiral 
$\gamma_5$ matrix in $2d$ dimensions so that the temperature dependent
part of the effective action could be decoupled through a
$\gamma_0$ rotation with constant phase. As
it is well-known, the resulting chiral Fujikawa Jacobian can be exactly
computed and yields to the $2d$ dimensional chiral anomaly.
This
gives another example of the connection
between C-S terms in odd-dimensions and even-dimensional topological
invariants connected to chiral anomalies \cite{J}-\cite{ADM}.

\noindent (iv) Although our result is obtained for a particular 
class of gauge field backgrounds (vanishing electric field
and time independent magnetic field in the Abelian case), similar
to those considered in the pioneering works at zero temperature
\cite{DJT}-\cite{R},  there is no doubt that the same gauge invariant
answer should be confirmed for general gauge-field configurations,
using for example a $\zeta$-function regularization analysis.

\noindent (v)  Remarkably, the temperature dependence of the parity-odd 
effective action is the same irrespectively of the number of space-time 
dimensions. This could be attributed to the particular background we 
considered but the topological nature of the result suggests that a similar 
result should hold in general. This is also sustained by the fact that the 
dependence on the $2d$-dimensional components of the gauge-field background 
occurs through the axial anomaly, also a quantity of topological nature.

We would like to end this work
by noting that the results we derived in a finite temperature 
Quantum Field Theory language, could also be interpreted in terms of a 
compactified Euclidean theory in an odd number of dimensions, where the 
curled coordinate is not necessarily the Euclidean time, but it may be a 
compact dimension of length $L = \beta$. If this interpretation is adopted, 
and one takes only the lowest Kaluza-Klein modes for the parity conserving 
part of the effective action, one then has a $2d$ reduced theory, where the 
odd part of the effective action we evaluated plays the role of a 
$\theta$-vacuum term (we assume, of course, that there is also a Yang-Mills 
action for the gauge field). 

~

\underline{Acknowledgements}: C.D.F. and G.L.R. are supported by CONICET, 
Argentina. F.A.S. is partially  suported by CICBA, Argentina and a 
Commission of the European Communities contract No:C11*-CT93-0315. This 
work is supported in part by funds provided by a CONICET grant PIP 4330/96 
and ANPCyT grants PICT 97 No:03-00000-02285, 03-00000-02249 and 
03-00000-0053.


\begin{references}
\bibitem{pis} R.~Pisarski, Phys.\ Rev.\ {\bf D35} (1987) 664.
\bibitem{bfs} N.~Brali\'c, C.D.~Fosco and F.A.~Schaposnik, Phys.\ Lett.\
{\bf B 383} (1996) 199.
\bibitem{cfrs} D.~Cabra, E.~Fradkin, G.L.~Rossini and F.A.~Schaposnik,
Phys.\ Lett.\ {B 383} (1996) 434.
\bibitem{dll} G.~Dunne, K.~Lee and Ch.~Lu, Phys.\ Rev.\ Lett.\
{\bf 78} (1997) 3434.
\bibitem{dgs} S.~Deser, L.~Griguolo and D.~Seminara, Phys.\ Rev.\ Lett.\
{\bf 79} (1997) 1976; Phys.\ Rev.\ {\bf D57} (1998) 7444.
\bibitem{frs} C.D.Fosco, G.L.Rossini and F.A.Schaposnik,
Phys.\ Rev.\ Lett.\ {\bf 79} (1997) 1980 and {\it errata ibid} {\bf 79}
(1997)
4296; Phys.\ Rev.\ {\bf D56} (1997) 6547.
\bibitem{JP1} R.~Jackiw and S.Y.~Pi, Phys. Lett. {\bf B423} (998) 364.
\bibitem{JP2}  R.~Jackiw and S.Y.~Pi, hep-th/9808036.
\bibitem{AF} I.J.R.Aitchinson and C.D.Fosco, Phys.\ Rev.\ {\bf D57}
(1998) 1171.
\bibitem{KN} Y.~Kikukawa and H.~Neuberger, Nucl. Phys. {\bf B513}
(1998) 735.
\bibitem{GF} R.~Gonz\'alez Felipe, Phys. Lett. {\bf B417} (1998)
114.
\bibitem{DD} A.~Das and G.~Dunne, Phys. Rev. {\bf D57} (1998) 5023
\bibitem{F} C.~Fosco, Phys. Rev. {\bf D57} (1998) 6554.
\bibitem{S} L.~Salcedo, hep-th/9802071.
\bibitem{BD} J.~Barcelos-Neto and A.~Das,
Phys. Rev. {\bf D58} (1998) 085022.
\bibitem{SSS} A.N.~Sisakian, O.Yu.~Shevchenko and S.B.~Solganik,
hep-th/9806047.
\bibitem{P} E.M.~Prodanov and S.~Sen, hep-th/9810044.
\bibitem{J} R.~Jackiw,Chern-Simons Terms and their Descendents in
Physical Theory, in Santa Fe 1984 proceedings, The Santa Fe Meeting, p.323.
\bibitem{ADM} L.~Alvarez-Gaum\'e, S.~Della Pietra and G.~Moore,
Ann. of Phys. (NY) {\bf 163} (1985) 288.
\bibitem{Ball} R.D.Ball, Phys.\ Rep.\ {\bf 182}, 1 \& 2 (1989).
\bibitem{wit} E.~Witten, On S Duality in Abelian Gauge Theory,
hep-th/9505186.
\bibitem{DJT} S.~Deser, R.~Jackiw and S.~Templeton,Phys. Rev. Lett. {\bf 48} 
(1982) 975; Ann. of Phys. (NY) {\bf 140} (1982) 372.
\bibitem{R} A.~Redlich, Phys. Rev. Lett. {\bf 52} (1982) 18;
Phys. Rev. {\bf D29} (1984) 2366.
\bibitem{GRS} R.E.~Gamboa Sarav\'\i ~, G.~Rossini and F.A.~Schaposnik,
Int. J. Mod. Phys. {\bf A11} (1996) 2643.
\end{references}
\end{document}